\documentclass[9pt,twocolumn,twoside]{opticajnl}
\journal{opticajournal} 

\setboolean{shortarticle}{true}

\usepackage{lipsum} 
\usepackage{xcolor}

\usepackage{lineno}

\title{Low dispersive phase-modulated rapid-scanning interferometry}

\author[1]{Fabian Richter}

\author[1]{Ulrich Bangert}

\author[1]{Friedemann Landmesser}

\author[1]{Nicolai Gölz}

\author[1]{Felix Riedel}

\author[1,*]{Lukas Bruder}

\affil[1]{Institute of Physics, University of Freiburg, Hermann-Herder-Str. 3, 79100 Freiburg, Germany}

\affil[*]{lukas.bruder@physik.uni-freiburg.de}

\begin{abstract}
Time-domain interferometry is an important principle in Fourier transform (FT) and nonlinear femto- to attosecond spectroscopy. 
To optimize the resolution and sensitivity of this approach, various interferometer stabilization schemes have been developed. 
Among them, acousto-optical phase modulation (AOPM) of the interferometer arms combined with phase-synchronous lock-in detection has proven as a particular sensitive technique. 
However, the acousto-optical modulators (AOMs), required for this technique, introduce several disadvantages. 
Here, we demonstrate an alternative phase modulation scheme which omits AOMs, termed PM scheme here. 
As a benchmark, we directly compare the performance between the PM and the AOPM scheme in a linear FT spectroscopy experiment and find comparable sensitivity in both approaches. 
\end{abstract}

\setboolean{displaycopyright}{false} 

\begin{document}

\maketitle

\section{Introduction}
Time-domain interferometry is the basic principle in FT spectroscopy\,\cite{smith_fundamentals_2011} as well as in femtosecond coherent multidimensional spectroscopy (CMDS) and related nonlinear spectroscopy schemes\,\cite{jonas_two-dimensional_2003, fuller_experimental_2015, bruder_efficient_2015}. 
However, performing these spectroscopy techniques in the optical regime requires a stabilization of the underlying interferometer. 
Moreover, nonlinear spectroscopy requires particular high sensitivity and high dynamic range to separate the weak nonlinear signals from the linear background\,\cite{fuller_experimental_2015}. 
The AOPM technique is an interferometer stabilization method which addresses these challenges very effectively\,\cite{tekavec_wave_2006}, as demonstrated in numerous applications\,\cite{lott_conformation_2011, bruder_coherent_2018, tiwari_spatially-resolved_2018, bruder_coherent_2019, wituschek_tracking_2020, autry_excitation_2020, uhl_improved_2022}. 
The method employs AOMs to imprint a rapid phase beating at frequency $\Omega$ between the interferometer arms ($\Omega$ denotes the frequency difference between the AOMs) which is combined with lock-in amplification. 
This shifts the signal from DC to a well-defined frequency $\Omega > 1$\,kHz sufficiently high to avoid the $1/f$-noise and other noise sources typically dominating in the frequency range $< 1\,$kHz\,\cite{bruder_phase-synchronous_2018}. 
At the same time, phase-synchronous lock-in detection with an optical reference removes most of the interferometer fluctuations from the signal. 
Thus, the AOPM method provides both particularly high sensitivity and a convenient passive interferometer stabilization. 

However, the use of AOMs in the optical path has several disadvantages. 
The AOMs introduce (i) optical losses, (ii) limit the spectral bandwidth acceptance of the interferometer, (iii) introduce significant temporal dispersion (and self-phase modulation in high-intensity applications), and (iv) introduce angular dispersion on the diffracted beams, detrimental for pulse compression and in microscopy applications.   
In principle, the bandwidth limitation can be overcome with super-continuum generation inside the interferometer\,\cite{javed_broadband_2024}, which is however limited to applications with low pulse energies. 
The angular dispersion may be solved with double-pass AOM geometries\,\cite{jana_fluorescence-detected_2024}, albeit at the expense of increased temporal dispersion. 
Alternatively, path-length modulation schemes have been demonstrated\,\cite{bloem_enhancing_2010, mirzajani_accurate_2022}, which however can introduce spurious nonlinearities. 
Likewise, related phase-cycling techniques based on acousto-optical and liquid-crystal pulse shapers\,\cite{tian_femtosecond_2003, vaughan_coherently_2007} exhibit similar disadvantages as the once mentioned above. 
Hence, a satisfying solution has not been developed, to our knowledge. 

In this letter, we introduce a shot-to-shot PM technique which omits AOMs (in the following termed PM technique). 
The method combines interferometer stabilization methods originally developed in FT spectroscopy\,\cite{wang_near_2008,wen_application_2023} with the signal-recovery advantages from the AOPM technique\,\cite{tekavec_wave_2006}. 
Key to our approach is rapid scanning of the interferometer, which is related to a recently demonstrated variant of the AOPM technique\,\cite{agathangelou_phase-modulated_2021}. 
For an assessment of the performance, we thus compare the PM technique with the rapid-scanning AOPM technique under identical experimental conditions. 
We find comparable signal-to-noise (SNR) performance for both schemes while our approach avoids the disadvantages of using AOMs.

\begin{figure}[ht]
	\centering
	\includegraphics[width=\linewidth]{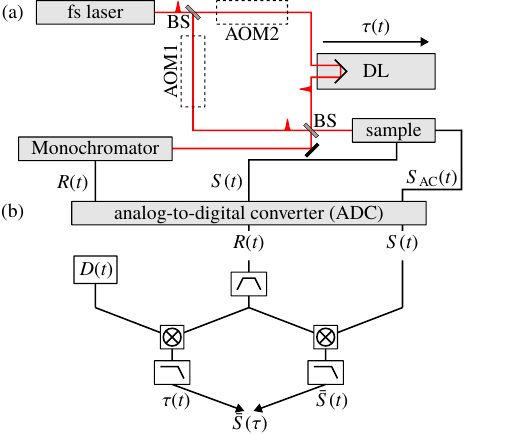}
	\caption{Experimental setup for the PM technique: optical setup (a), signal processing scheme (b). AOMs, shown as dashed boxes, are only installed for the AOPM technique. 
    The delay line (DL) is continuously scanned introducing a time shift $\tau(t)$ in the interferometer arm. 
    The signals $S(t)$, $R(t)$ and $S_\mathrm{AC}(t)$ are simultaneously digitized with a multi-channel analog-to-digital converter (ADC), whereas $D(t)$ is directly generated in the computer. $\bar{S}(t)$ and $\bar{S}(\tau)$ denote the demodulated signal before/after mapping the laboratory time $t$ onto the interferometer delay $\tau$. 
    Further details in the main text.}
	\label{fig:setup_analysis}
\end{figure}

\section{Experimental setup and data processing}
The experimental setup used for the new PM technique along with the signal processing scheme is shown in Fig.\,\ref{fig:setup_analysis}. 
The same optical setup is used for the AOPM measurements with the only difference that we install an AOM in each interferometer arm and slightly modify the data processing scheme\,\cite{landmesser_high-resolution_2025}. 
The AOPM technique, including the rapid-scanning variant, has been described in detail before\,\cite{tekavec_wave_2006,agathangelou_phase-modulated_2021}. 
We thus focus our description on the PM scheme in the following. 

A Mach-Zehnder interferometer equipped with a mechanical delay line (DL) is used to create a pair of collinear pulses from a femtosecond (fs) laser system (80 MHz repetition rate, 200 fs pulse duration). 
One part of the interferometer output excites a vapor cell containing a mixture of K, Rb and Cs vapor (kept at room temperature). 
The sharp, well-defined D-lines of alkali atoms serve us as benchmark to characterize the technique. 
The fluorescence $S$ is detected as a function of the optical delay $\tau$ between the interferometer arms. 
A Fourier transform of $S(\tau)$ yields the light-induced fluorescence spectrum. 
Hence, the principle is based on FT spectroscopy. 
However, instead of measuring the optical absorption in a sample, action detection is used. 
Action signals can be fluorescence, as in the presented case, or photocurrent and photoionization yields\,\cite{tekavec_wave_2006, karki_coherent_2014, bruder_coherent_2018, uhl_coherent_2021}. 

We apply rapid, continuous scanning of the delay line in the interferometer at a velocity of $v$. 
This leads to a phase modulation of the interference signal $S$ at the Doppler frequency
\begin{equation}
\omega_D = \frac{2 n v  \omega}{c}\, ,
\label{eq:doppler_freq}
\end{equation}
where $n$ denotes the refractive index of air and $c$ the speed of light in vacuum, and $\omega$ the optical frequency. 
For reasonable experimental parameters (see below) we get $\omega_D = 1.9$\,kHz $> 1\,$kHz. 
We thus effectively shift the signal $S$ to a modulation frequency sufficiently high to avoid the low-frequency noise in the laboratory. 
Even higher modulation frequencies may be readily achieved with optimized delay lines or in experiments at shorter wavelengths in the visible to ultraviolet spectral range. 

The $\omega_D$-modulation is used for phase-synchronous lock-in amplification (Fig.\,\ref{fig:setup_analysis}b). 
To this end, a portion of the interferometer output is spectrally filtered in a monochromator (center frequency $\omega_R$) yielding the signal $R$. 
This signal essentially tracks the phase evolution of the interferometer evaluated at the frequency $\omega_R$. 
$S$ and $R$ are synchronously digitized (sampling rate of 100\,kSa/s). 
Further signal processing is performed on a personal computer. 
$R(t)$ is band-pass filtered at $\omega_D$ (Butterworth filter, 3dB-bandwidth: 350\,Hz) and Hilbert-transformed to construct a complex-valued waveform with normalized amplitude. 
$R$ and $S$ are multiplied and subsequently low-pass filtered (3dB-bandwidth: 350\,Hz) yielding the demodulated signal $\bar{S}$. 
The differential phase measurement between $S$ and $R$ strongly reduces the residual phase noise in $\bar{S}$. 
This lock-in detection scheme has thus two advantages: 
The reduction of the phase noise leads to an effective passive interferometer stabilization, enabling FT interferometry even in the extreme ultraviolet spectral domain\,\cite{wituschek_tracking_2020, uhl_improved_2022}.
Moreover the lock-in detection greatly increases the detection sensitivity, as it is generally known from lock-in amplification. 

In addition, we use the $\omega_D$-modulation for real-time tracking of the time-axis $\tau(t)$ of the interferogram $\bar{S}$. 
In a separate computing loop, a sine-wave $D(t)$ at frequency $\omega_0$ (chosen close to $\omega_D$) is generated to demodulate $R(t)$ analogous to the procedure described above. 
$\tau(t)$ is then reconstructed from the unwrapped phase of the demodulated signal by using the known frequencies $\omega_R$ and $\omega_0$. 
This determines relative changes in $\tau(t)$ in real time and thus compensates for drifts in the scan velocity $v$ and other phase jitter picked up in the interferometer. 
For the calibration of the absolute delay, we determine $\tau(t)=0\,$fs from the linear autocorrelation signal $S_\mathrm{AC}$ between the two interferometer arms, recorded while scanning the delay line. 
Note, that for the described procedure $v$ does not need to be constant. However, close to the turning points of the delay line, the modulation frequency $\omega_D$ becomes unfavorably small which is why we discard this part of the data. 
Eventually, $\tau$ and $\bar{S}$ are binned to equidistant $\tau$-steps of 1\,fs. 
We apply zero padding and a Gaussian window function followed by a Fast Fourier Transform (FFT) of $S$ with respect to $\tau$. 
With this procedure, the real-part of the FFT spectrum corresponds to the linear excitation spectrum and the imaginary-part to the dispersion of the sample\,\cite{tekavec_wave_2006}. 

The described signal processing steps are similar to the AOPM technique\,\cite{tekavec_wave_2006, agathangelou_phase-modulated_2021}. 
However, in the latter, AOMs are used to induce a well-defined, very sharp phase modulation $\Omega$, that is decoupled from the interferometer delay $\tau$. 
This provides ideal conditions for efficient signal filtering and the tracking of the interferometer delay. 
In contrast, the $\omega_D$-modulation in the PM technique is significantly broadened by the width of the femtosecond laser spectrum and fluctuations in the delay stage velocity $v$. 
This may have an effect on the signal processing methods and their performance. 
Yet, for reasonable laser parameters (Fourier limit $> 10\,$fs, wavelength of 800\,nm) $\Delta \omega_D / \omega_D < 0.1$ which is sufficiently narrow for efficient signal filtering. 

\section{Results}
\begin{figure}[t]
	\centering
	\includegraphics[width=\linewidth]{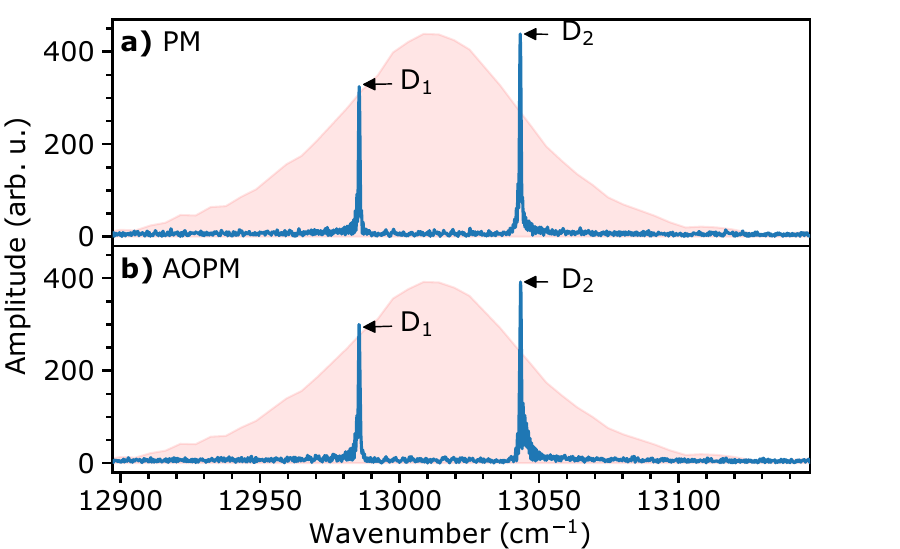}
	\caption{Fluorescence-detected excitation spectrum of K atoms recorded with the PM (a) and with the AOPM (b) technique. Shown is the absolute-value of the Fourier spectrum. Labels D1, D2 denote the D-line resonances of the K atoms. Red envelopes show the used laser spectrum, centered at 769\,nm.}
	\label{fig:K_comparison}
\end{figure}
The PM and AOPM techniques are compared with systematic measurements of the D-line excitations in K and Rb. 
An example FFT spectrum is shown in Fig.\,\ref{fig:K_comparison}. 
Here, we show the absolute value of the FFT for convenience. 
For both, the PM and AOPM measurement, the delay line was scanned in the range of -20\,ps to 60\,ps at a nominal stage velocity of 1.49\, mm/s (10\,ps/s), corresponding to a modulation frequency of $\omega_D =$1.9\,kHz. 
In the AOPM case, the AOM-induced modulation was $\Omega = 5$\,kHz, hence the total modulation frequency was here $\Omega + \omega_D = 6.9$\,kHz. 
To extract the modulated signals, a bandpass filter was centered at the respective frequency at $\omega_D$ (in the PM case) and at $\Omega + \omega_D$ (in the AOPM case) (Fig.\,\ref{fig:setup_analysis}b). 
For the signals $R$ and $S$, only the delay range of 0.2 to 58\,ps was considered. 
This avoids the delay range with strong acceleration/deceleration and the pulse-overlap range where stray light significantly compromises the interferogram due to the low fluorescence yield. 
Note, that in principle less data need to be spared, however, for convenience reasons we cut the data fairly far away from the acceleration/deceleration regions. 
No averaging was applied, only a single delay scan was Fourier transformed to get the spectra in Fig.\,\ref{fig:K_comparison}. 
For the evaluation of $S_\mathrm{AC}$, we demodulate the full signal range and determine the zero delay position with a Gaussian fit. 

The PM and AOPM measurements exhibit both very good quality apparent from the large signal-to-noise ratio (SNR) of the Fourier spectra. 
Residual delay inaccuracies lead to minor ghosting which causes the slight broadening and asymmetry of the peak  pedestals. 
This effect is actually slightly larger in the AOPM measurement. 
The slight difference in the peak amplitudes are within the statistical uncertainty of the experiment (see below). 

\begin{figure}[t]
	\centering
	\includegraphics[width=\linewidth]{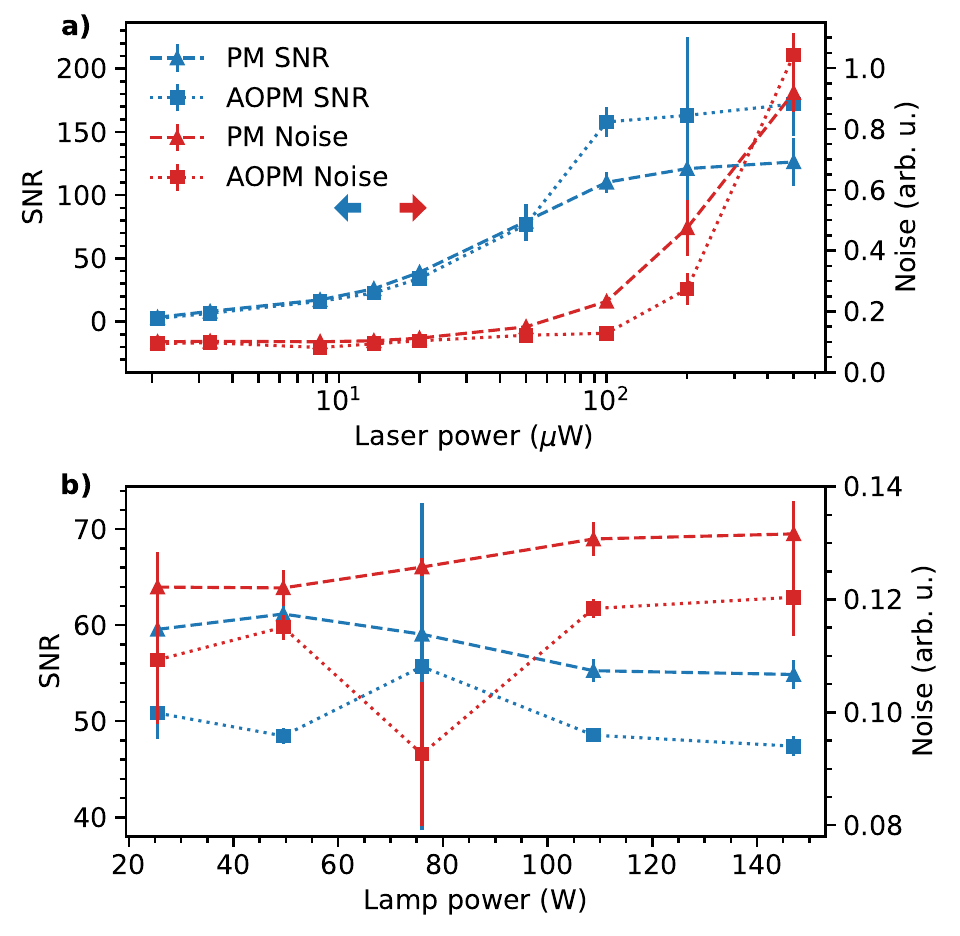}
	\caption{Performance comparison. SNR (left scale) and noise level (right scale) for two different measurement scenarios: (a) attenuation of laser power exciting the sample, (b) introduction of phase fluctuations in the interferometer with a heat source. As heat source served a halogen lamp, whose electrical power is given on the x-axis.}
	\label{fig:Power_attenuation}
\end{figure}
For a more quantitative comparison we repeated essentially the same measurement for Rb under a broad range of different SNR conditions. 
Here, the laser spectrum was centered at the D2 line of Rb (12820\,cm$^{-1}$) to minimize fluctuations in the fluorescence yield due to drifts in the laser spectrum and to avoid the excitation of other resonance in the sample.
Specifically, we investigated two scenarios: 
(i) We reduced the laser power exciting the sample. 
This deteriorates only the detected fluorescence signal $S$, while the quality of the reference signal $R$ is not affected. 
Hence, we model here the issue of measuring very weak signals from a sample. 
(ii) We artificially introduced phase fluctuations in the interferometer by placing a heat source (halogen lamp) at a distance of 5\,mm to the beam path of one interferometer arm. 
This causes fluctuations of the refractive index and thus of the phase between the two interferometer arms. 
The electrical power applied to the halogen lamp controls the temperature (varied in the range $21^\circ$C to 165$^\circ$C) and thus the magnitude of the refractive index fluctuations in the surrounding air. 
Note, that method (ii) affects both $S$ and $R$ and thus not only deteriorates the fluorescence signal but also the ability to track and correct the delay/phase fluctuations in the interferometer. 
Hence, this case models measurements in noisy environments. 

The SNR of the Fourier spectra was computed as the ratio between the amplitude of the D2 line divided by the root-mean-square (RMS) value of the noise floor. 
The latter was evaluated in the spectral range 12586 to 12786\,cm$^{-1}$. 
Error bars indicate the deviation between 2 consecutive measurements.  
Fig.\,\ref{fig:Power_attenuation} shows the evolution of the SNR for both scenarios. 
Additionally, we show separately the noise floor for each measurement.  
For comparison the example data shown in Fig.\,\ref{fig:K_comparison} exhibits a SNR of $\approx 90$. 

In case (i), the SNR is varied over a large range. 
Both the SNR and the noise level increase with increasing laser power (Fig.\,\ref{fig:Power_attenuation}a). 
This is expected since the signal amplitude scales linear and the noise level with the square root of the laser power. 
The atomic transition starts to saturate at a laser power of $> 50$\,µW. 
This coincides with an increased noise level in the PM method compared to the AOPM method. 
A closer inspection reveals, that this is a problem of the smaller modulation frequency used in the PM method and, hence, can in principle be avoided. 
In case (ii), the changes in SNR and noise level are marginal (Fig.\,\ref{fig:Power_attenuation}b). 
Only a small increase in the noise level accompanied by a small decrease of the SNR is observed. 
This is expected, since the phase fluctuations in the interferometer do not alter the laser power or spectrum and thus are expected to affect only the noise floor of the signal but not the signal amplitude. 
As an important result, under clean (non saturation) conditions the PM and AOPM method deliver in both cases comparable performance. 

We also investigated the effect of the filter bandwidths implemented in the signal processing scheme. 
Since the PM and AOPM methods employ different modulation schemes, we would have expected an influence of this parameter on the overall performance. 
However, we could not detect a significant difference in the SNR when altering the bandwidths from 25\,Hz to 1\,kHz. 
We note, that in rapid-scanning operation some delay stages introduce periodic oscillations in the scan speed. 
In our case (Newport DL125 stage) this lead to a pronounced satellite peak in the FFT spectra which is slightly stronger in the PM method compared to the AOPM method (not shown). 
However, the artifact disappears at high scan speeds (factor of 7 higher scan speed than used here). 

\begin{figure}[ht]
\centering
\includegraphics[width=\linewidth]{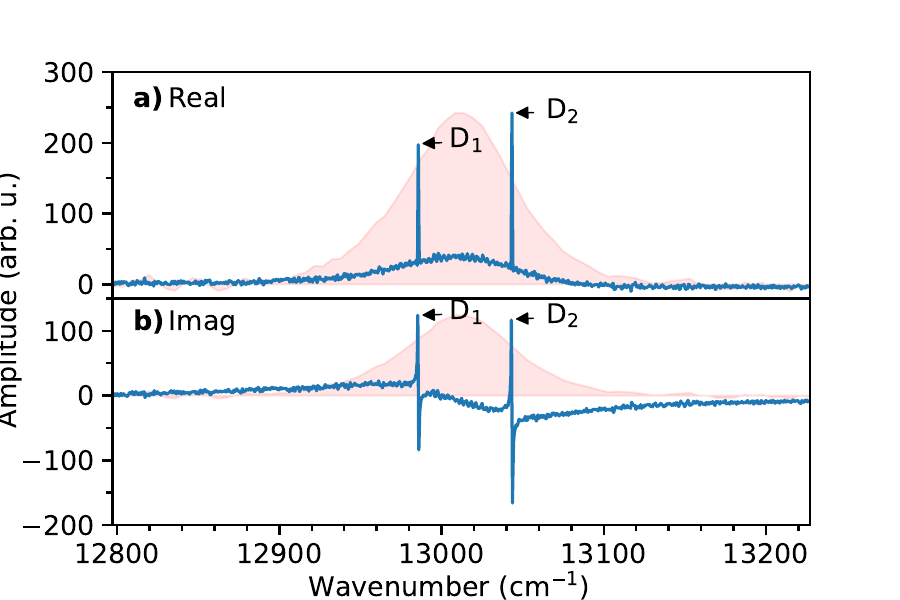}
\caption{Decomposition of the Fourier spectrum into absorptive (a) and dispersive lineshape (b). 
}
\label{fig:Phasing}
\end{figure}
So far we have evaluated the absolute value of the Fourier spectra. 
However, the phase-synchronous demodulation scheme allows for a decomposition into the absorptive (real part of $S$) and the dispersive part (imaginary part of $S$) of the spectrum without relying on the Kramers-Kronig relation\,\cite{tekavec_wave_2006}. 
This is advantageous in nonlinear spectroscopy where the Kramers-Kronig relation does not necessarily hold. 
However, a correct separation of the absorptive and dispersive line shape requires the elimination of phase errors in the time-domain interferogram. 
For this purpose, the zero delay point $\tau= 0\,$fs has to be precisely determined, which we extract from the linear autocorrelation measurement included in our signal processing scheme. 
In addition, the global phase offset between $S$ and $R$ has to be determined and taken into account\,\cite{tekavec_wave_2006}. 
Fig.\,\ref{fig:Phasing} confirms that the correct separation of the absorptive and dispersive contribution is possible with the PM method. 
Here, we evaluated the data for a delay range of 0 - 58\,ps. 
Due to the low vapor density and thus low fluorescence yields, we observe a significant contribution from stray light (broad peak centered at 13010 cm$^{-1}$) which, however, shows also the correct decomposition of the signal contributions. 

\section{Conclusion}
We introduced a new scheme for rapid-scanning phase-modulated interferometry. 
A direct comparison between our PM technique and the previous AOPM technique, tested under a range of different noise conditions, confirms comparable performance for both methods. 
Yet, our method omits the use of AOMs or other phase-shaping devices, which is advantageous in view of optical losses, spectral bandwidth, temporal and angular dispersion and makes the technique compact and easy to implement. 
The optical setup is very simple and features particularly low dispersion (in total 1.2\,mm of fused silica is sufficient), which is a factor of $\geq 10$ smaller than in AOPM setups. 

While the signal processing was performed here with a custom algorithm on a personal computer, the method can be also implemented with commercially available lock-in amplifiers. 
Potential applications of our technique would be in FT spectroscopy to increase the general sensitivity or to implement action detection. 
The technique may be also used in coherently-detected two-dimensional spectroscopy performed in the pump-probe geometry\,\cite{fuller_experimental_2015} or even in action-detected two-dimensional spectroscopy. 
The latter could be implemented in analogy to Ref.\,\cite{lomsadze_tri-comb_2018}.


\begin{backmatter}

\bmsection{Acknowledgment} The following funding is acknowledged: European Research Council (ERC) Starting Grant MULTIPLEX (101078689), Deutsche Forschungsgemeinschaft (DFG) RTG 2717, Baden-Württemberg Stiftung Eliteprogram for Postdocs.

\bmsection{Disclosures} The authors declare no conflicts of interest.

\bmsection{Data Availability Statement} The experimental data included in this work are available on the open repository: \textit{Accession codes will be available before publication.}

\end{backmatter}

\bibliography{paper_bib_file,2025_Rapid_Scan_noAOM}

\bibliographyfullrefs{paper_bib_file,2025_Rapid_Scan_noAOM}


\end{document}